# Performance of the first prototype of the HAWC Gamma Ray Observatory

Andres Sandoval, Rubén Alfaro, Ernesto Belmont, Varlen Grabski, Alejandro Rentería, Alejandro R. Vásques, Omar Vázquez
*Instituto de Fisica, UNAM, Mexico*

M. Magdalena González
*Instituto de Astronomia, UNAM, Mexico*

Alberto Carramiñana

Instituto Nacional de Astrofisica Electronica y Optica, INAOE, Mexico

Cesar Alvarez
*CEFYMAP-Universidad Autonoma de Chiapas, Mexico*

For the HAWC Collaboration

The HAWC gamma ray observatory, to be constructed at Sierra Negra, Puebla in Mexico, is a large array of water Cherenkov detectors sited at an elevation of 4100 m, which has been optimized for gamma/hadron discrimination of the primary cosmic rays in the TeV energy range. It is based on the Milagro experience, but the design has been changed from a water pond to individual water tanks. In order to validate the design with large water tanks a prototype array has been constructed near the HAWC site with 3 of the largest commercial rotomolded plastic tanks available in Mexico. They have been instrumented with 8" hemispherical photomultiplier tubes and read out with 2 Gsample/s flash ADCs. The performance of a single tank has been measured as well as the response of the array to cosmic ray showers. In this paper we present the first measurements of the performance of the HAWC prototype array.

## 1. INTRODUCTION

The High Altitude Water Cherenkov observatory (HAWC) [1] is a second-generation gamma ray detector that will survey the sky continuously from it site at 4100 m altitude and $19^o$ northern latitude at the Sierra Negra volcano in the state of Puebla, 200 km in a straight line from Mexico City. The baseline detector design called for 900 Cherenkov detectors in large water tanks (4m diameter x 5m depth) each instrumented with a single 20 cm photomultiplier tube looking up at the water volume from the bottom of the tank. The tanks will be densely packed to cover an area of ~25,000 m2. The array detects atmospheric cascades through the Cerenkov light produced in the water of the tanks and has a good gamma/hadron discrimination of the primary cosmic ray in the energy range of hundreds of GeV to 100 TeV and an angular resolution between $0.5^o$ and $0.25^o$. HAWC is designed to look for GRBs, monitor AGN for variability and measure the high energy behavior of galactic gamma-ray sources.

To validate the conceptual design, face the reality of commercial vs. custom built water tanks, prove the strength and commitment of the US-Mexican collaboration and evaluate the performance of a small array it was decided to build a prototype consisting of 3 Cherenkov detectors using large commercial water tanks in an environment like the one of the final observatory.

## 2. THE PROTOTYPE ARRAY

As the HAWC site at 4100m above sea level is being developed it was decided to place the prototype array higher up in the mountain at 4530m close to the top of Sierra Negra, where the Large Millimetric Telescope (LMT) has been built, a radio telescope with a 50m antenna. The LMT collaboration has developed the site, which counts with road access, electricity, Internet, security and teams of engineers and skilled workers going daily up the mountain.

In Mexico every house has a rotomolded plastic water tank on the roof. The largest manufacturer of these water tanks is ROTOPLAS. This company was approached by the Auger collaboration to design their prototype tanks and when the final design was chosen, it built a tank factory in Argentina to produce a large fraction of the 1600 tanks of the Auger array.

With funds from the Physics Institute at UNAM and the University of Chiapas we bought 3 of the largest water tanks manufactured by ROTOPLAS, of 3m diameter and 3.6m height, which were delivered between November and December 2008. To manufacture 4m diameter and 5m tall tanks, as the ones of the base design, the company needs to build a larger oven than the largest available at present in Mexico as well as a new mold. With such large tanks we would face the problem of transportation, as they exceed the dimensions of what is allowed on the Mexican highways. All this indicates that 4m diameter plastic tanks transported from a factory in Northern Mexico might not be viable. Therefore a tank factory near the site would be needed. At present several Mexican and US companies have made a bid to set up a tank production chain in Ciudad Serdan, 20 Km from the HAWC site.

There exists a second option to the plastic tanks namely metallic pipes of 7.3m diameter and 5m height with an inner light-tight bag. This solution is being tested in California and Colorado by the collaboration.

To make the tanks completely light tight the company had to mix 2% of very fine carbon powder in the raw material before rotomolding it, making sure it is evenly dispersed and not clumped. They also increase the UV resistance of the plastic to be able to sustain the increased UV levels at the top of the mountain.

In five field trips to the mountain between November 2008 and March 2009, for a total of 14 days, we cleaned and setup the 3 tanks, filled them with filtered and softened water and instrumented them with photomultipliers that have been recycled from the Milagro gamma ray observatory (Fig.1).

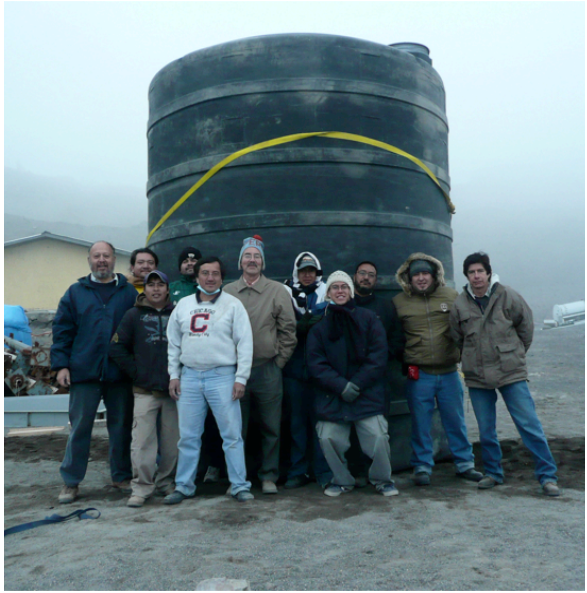

Figure 1. The first tank of the HAWC prototype Cherenkov detector in the volcano Sierra Negra, Mexico at an altitude of 4530m.

The detectors have been setup in a triangular pattern (Fig.2) and named T1, T2 and T3. Table 1 gives their coordinates.

Table 1: Coordinates of the centers of the 3 Cherenkov detectors.

|    | x | y | z |
|----|-----|-----|-----|
| T1 | 0.0 m | 0.0 m | 4531.29 m |
| T2 | -14.59 m | -27.76 m | 4531.12 m |
| T3 | -33.19 m | 0.60 m | 4531.05 m |

Each tank holds 25,000 l of water, which had to be carried up from a well at the bottom of the mountain by trucks. To supply the water for the HAWC array a well will be dug near the site. The tanks were filled with filtered and softened water. A manhole of 80cm diameter allows access to the interior of the tank .

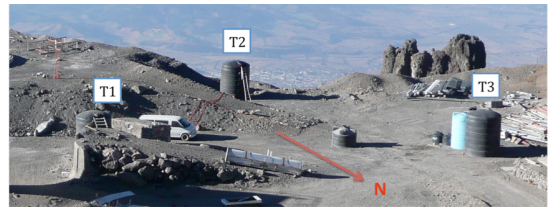

Figure 2. The 3 water Cherenkov detectors of the HAWC prototype.

The contractors building the LMT to put a concrete plant for the structure supporting the antenna had leveled the site of roughly 80x80 m2. It contains still some leftover structures like a ramp near the T1 tank and a small stone warehouse also next to T1 that served us as a counting room.

The T1 tank was instrumented with three 8″ Hamamatsu R5912 photomultipliers that had been used in Milagro [2]. The voltage divider of the PMTs is encased in a watertight plastic cylinder from which emerges the high voltage cable carrying also the signal. One PMT was placed looking down on the top of the water (T), a second (C) was placed centered at the bottom of the tank and the third (S) sat at the side close to the tank wall, see Fig.3. Two scintillator paddles 60 x 60 cm2 were used to select vertical muons by being placed one at the top and the other at the bottom of the tank.

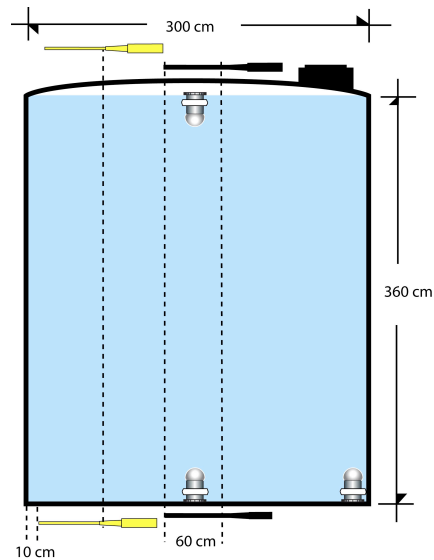

Figure 3. Layout of the T1 detector with 3 PMTs, one on the top (T) floating and looking down, two on the bottom one centered (C) and one on the side of the tank (S). Two scintillator paddles (P1, P2) with 60x60 cm2 area were used to trigger on vertical muons at two different positions.

The other two tanks, T2 and T3, were instrumented with only one fotomultiplier centered at the bottom of the tank.

The HAWC array is optimized to obtain the best timing resolution for the arrival time of the signals rather than to collect the maximum amount of the Cherenkov light produced in each tank. Therefore the walls and bottom of the tanks are absorbing and not reflecting, as is the case in the Auger tanks. The signals are consequently faster but of lower amplitude.

## 3. MEASUREMENTS

The photomultiplier signals were split by a linear fan out with one of them fed to a discriminator and a gate generator to make the trigger signal, the other went into a 4 channel CAEN V1729 FADC that digitized the signal shape taking up to 2 Gsamples/s. Offline analysis provided the amplitude, integrated charge and arrival time of the Cherenkov light produced by the incoming particles as seen by the fotomultiplier. Typical pulse shapes for air showers detected by the three detectors in coincidence are shown in Fig. 4. The time difference between the arrival times of the signals in the three tanks gives the direction of the shower, permitting to obtain a rough sky map of cosmic rays.

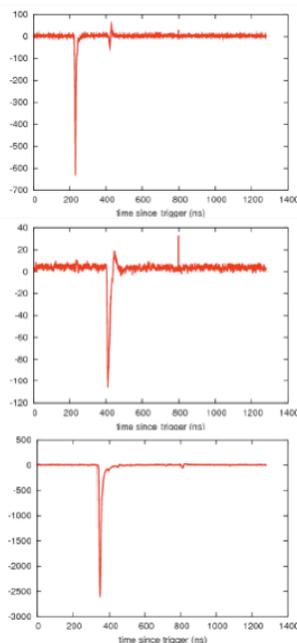

Figure 4. Pulse shapes from the three Cherenkov detectors for a shower digitized with a Flash ADC.

The count rate of signals detected in a 3m diameter water tank by the photomultiplier at the center was of the order of 16 KHz with the trigger set at a fraction of a single photo electron.

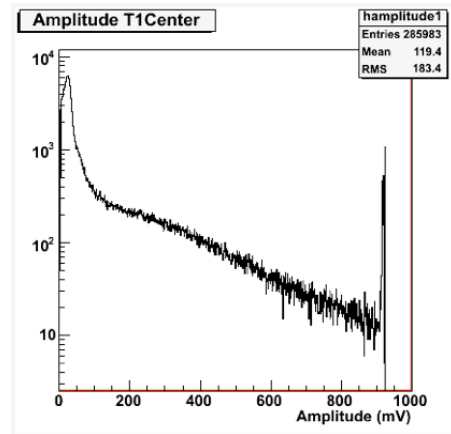

Figure 5. Pulse height spectrum of the water Cherenkov detector T1.

Amplitude spectra for one single tank are shown in Fig. 5 for (C), the central photomultiplier in T1. They correspond to events triggered by the C counter, and to vertical muons selected with the scintillator paddles centred on the tank and for muons selected with the scintillator paddles positioned 1m from the centre of the tank. One can see the vertical muon peak when triggering on the paddles and how it moves with the distance of the muon to the PMT.

In order to use the array to trigger on air showers, the output of the discriminators were broadened to 250 ns and a triple coincidence was required. The resulting trigger rate was 1 Hz. The amplitude spectra and distribution of the arrival time differences between the detectors was analyzed. From the arrival times of the shower particles at the three tanks the direction of the primary cosmic ray was reconstructed and the resulting sky map is shown in Fig. 6.

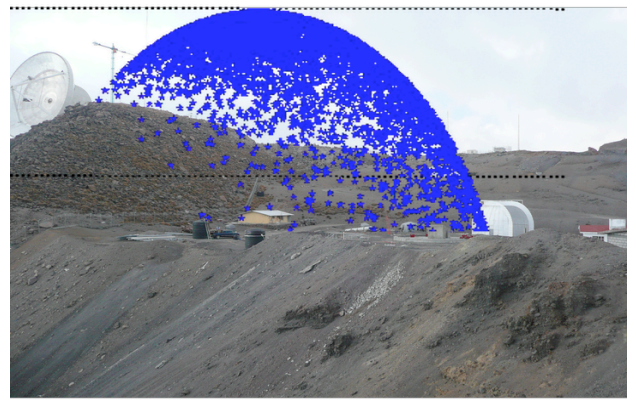

Figure 6. Sky map of primary cosmic rays seen by the water Cherenkov detectors of the HAWC prototype.

## 4. CONCLUSIONS

A prototype of the HAWC gamma ray observatory has been constructed and made operational near the chosen site at the Sierra Negra volcano in Mexico, at 4320 m altitude. The original HAWC design consists of 900 of water Cherenkov detectors optimized for gamma/hadron discrimination and sub-nanosecond timing for reconstruction of the shower direction with good accuracy. This calls for large water tanks of 4m diameter and 5m height for the Cherenkov detectors. As these are not commercially available, the prototype array was made with three rotomolded plastic tanks of 3m diameter and 3.6m height. They were filled with purified water and instrumented with 8" Hamamatsu R-59312 photomultipliers able to detect single photons.

Data with a single tank was obtained with a trigger PMT at the bottom center of the tank. Air showers were recorded by the triple coincidence of the tank signals within a 250ns time window. From the arrival time differences of the signals in the three tanks the direction of the shower was reconstructed and a sky map of the primary cosmic rays was produced.

To make the 900 polyethylene water tanks of the base design a new factory has to be setup near the site with larger ovens a new mold.

Another option looks more promising consists of making only 300 Cherenkov detectors in larger water tanks of 7.3m diameter and instrumented with three photomultipliers each. A prototype water tank using a cylinder of galvanized steel sheets assembled on site with a water- and light-tight bag has been tested as shown in Figure 7.

This exercise amalgamated the collaboration between Mexican and US groups and demonstrated that the infrastructure exists at the Sierra Negra site to be able to build the HAWC gamma ray observatory. The next stage is to construct an array of 6 large metallic tanks, 7.3m diameter at the site of the HAWC observatory.

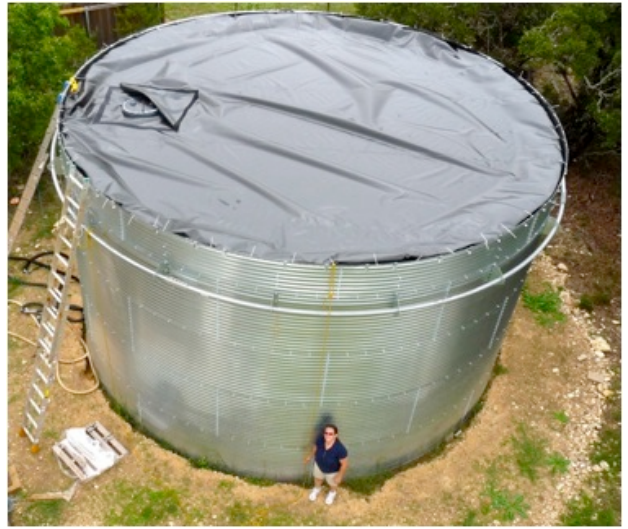

Figure 7. Test of a 7.3m diameter water tank.

## 5. ACKNOWLEDGEMENTS

The authors acknowledge the support by Direccion General de Apoyo al Personal Academico (DGAPA), UNAM under projects IN15507 and IN119708 to Consejo Nacional de Ciencia y Tecnología (CONACYT) under project 57674 , Instituto de Astronomia under project CI02 and Promep under project PROMEP/103.5/08/3291.